\documentstyle[pra,twocolumn,aps]{revtex}

   \input epsf
   \begin{document}
   \draft
   \title{Free expansion of a Bose-Einstein condensate in a 1D optical
lattice}
   \author{O.~Morsch, M.~Cristiani, J.H.~M\"uller, D.~Ciampini,
   and E.~Arimondo}
   \address{INFM, Dipartimento di Fisica E.Fermi, Universit\`{a} di Pisa, Via
   Buonarroti 2, I-56127 Pisa, Italy}
   \date{\today}
   \maketitle
   \begin{abstract}
   We have experimentally investigated the free expansion of a
   Bose-Einstein condensate in an array of two-dimensional traps
   created by a one-dimensional optical lattice. If the condensate
   held in a magnetic trap is loaded adiabatically into the lattice, the
   increase in chemical potential due to the additional periodic
   potential is reflected in the expansion of the condensate after
   switching off the magnetic trap. We have calculated the chemical
   potential from measurements of the transverse expansion of the condensate as a function of
   the lattice
   parameters.
   \end{abstract}
 The properties of Bose-Einstein condensates (BECs) in
lower-dimensional
   trapping potentials have recently attracted increasing interest.
   Both theoretical and experimental investigations have been aimed
   at studying the differences between 1D~\cite{bongs01,schreck01,greiner01,gorlitz01,plaja02,salasnich02,kramer02}
   and 2D
   traps~\cite{anderson98,javanainen99,taichenachev99,petrov00,prokofev01,pedri01,burger02,gorlitz01,salasnich02,kramer02}
   with regard to the static and dynamic behaviour of BECs in such
   potentials.
   A thorough understanding of these properties is
   important as a prerequisite for predicting, e.g., the evolution of a
   BEC loaded into 1D and 2D wave-guides. Condensates in 2D and 1D
   have been realized in magnetic traps starting from a 3D situation
by changing the aspect
   ratio of the trap and the number of atoms in the
condensate~\cite{gorlitz01}.
   In order to realize a 1D condensate, optical dipole traps have
   been used to achieve the necessary asymmetry between the trapping
   frequencies~\cite{schreck01,greiner01}. Similarly, 2D condensates can be created in an
   array of pancake-shaped traps provided by the periodic potential of
a 1D optical
   lattice~\cite{anderson98,burger02}. For large lattice depths, i.e. in the
   tight-binding regime, the trapping
   frequencies along the lattice direction exceed those of the
   magnetic trap by orders of magnitude and the 1D lattice hence
   represents an array of 2D traps in which the motion along the
   direction of the array is frozen. Recently, Pedri {\em et
   al.}~\cite{pedri01} have
   calculated the variation in chemical potential when the depth of
   the periodic potential is increased.

   In this paper, we report
   on experiments with magnetically trapped BECs loaded into a
one-dimensional optical
   lattice and subsequently allowed to expand freely inside the
   lattice.
   After the creation of an array of 2D traps within the 1D
   optical lattice, the magnetic trap is
   suddenly switched off. The condensate is then free to expand in the
   radial directions whilst still being confined in the lattice
   direction. The measured expansion allowed us to infer the chemical
   potential and to test its dependence on the lattice parameters.

   Our experimental setup for creating Bose-Einstein condensates of
   rubidium atoms and loading these into 1D optical lattices has
   been described in detail elsewhere~\cite{muller00,morsch01}.
Briefly, we
   obtain BECs of $\approx 1-3\times 10^4$ rubidium atoms inside a
triaxial TOP-trap
   which are then loaded into a 1D optical lattice created by two
   independently controllable, linearly polarized Gaussian laser beams
of
   wave-vector $k$ and waist $w=1.8\,\mathrm{mm}$, detuned by
$25-40\,\mathrm{GHz}$ from the rubidium resonance line
   and propagating at an angle $\theta$ with respect to each
   other. The lattice constant $d$ of the resulting one-dimensional
   periodic potential is then given by
  $d = \pi/(k\sin(\theta/2)),$
   and in the following we shall express the lattice depth $U_0$ in
units
   of a re-scaled lattice recoil energy $E_{rec}=\hbar^2
   \pi^2/(2md^2)$~\cite{morsch01} corresponding to the respective
lattice spacing $d$. In our setup we could realize both
   a horizontal lattice with $d=1.56\,\mathrm{\mu m}$ and a
   vertical lattice with $d=1.2\,\mathrm{\mu m}$. In the vertical case, after
   switching off the magnetic trap we accelerated the lattice downwards with an
acceleration
   $a=9.81\,\mathrm{m\,s^{-2}}$ by chirping
    the frequency difference between the lattice beams in order to
compensate gravity in the rest frame of the lattice. Apart from
the different lattice spacing, the vertical and horizontal cases
were, therefore, equivalent.

   \begin{figure}
   \centering\begin{center}\mbox{\epsfxsize 2.8 in \epsfbox{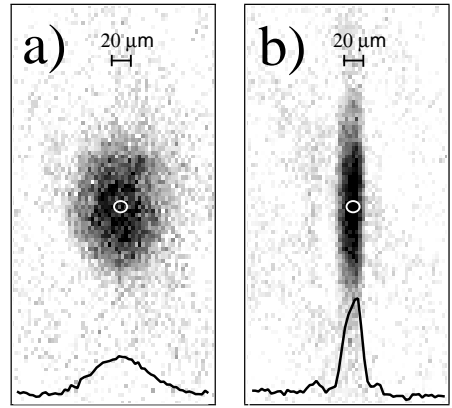}}
   \caption{Expanded condensate after $23\,\mathrm{ms}$ of
time-of-flight without a lattice (a) and with
   a lattice of depth $U_0=20\,E_{rec}$ present during the
   expansion (b). The line plots represent the integrated profiles of
   the condensate density along the lattice direction; the reduced width $\rho_{||}$ along that
   direction (horizontal in this figure) is clearly
visible. The white ellipses indicate the initial size of the
condensate in the trap.}\label{Fig1:withandwithout}
   \end{center}\end{figure}

\begin{figure}
   \centering\begin{center}\mbox{\epsfxsize 2.8 in \epsfbox{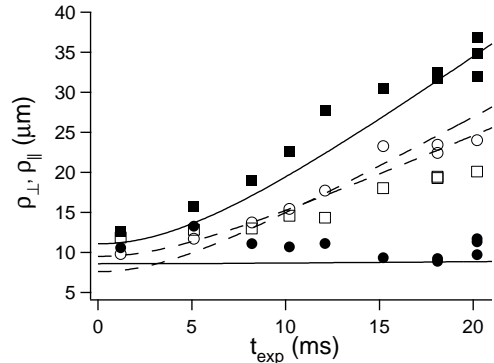}}
   \caption{Condensate dimensions (single shot values) $\rho_{||}$ (circles) and
   $\rho_{\perp}$ (squares), versus the expansion time $t_{exp}$ in
the presence of a horizontal lattice of
   depth $U_0\approx 25\,E_{rec}$ (filled symbols) and without lattice
(open symbols). The condensate was initially released from a trap
with $\overline{\omega}_{trap}/2\pi=25\,\mathrm{Hz}$
($\omega_{\perp}/\omega_{||}=1/\sqrt{2}$). The dashed and solid
lines are the results of a numerical simulation (see
text).}\label{Fig2:expansion}
   \end{center}\end{figure}

   In order to load the condensate into the optical lattice, after
   adiabatically reducing the mean magnetic trap frequency
${\overline{\omega}}_{trap}$ to the
   desired value we linearly increased the lattice depth from $0$
   to $U_0$ in a time $t_{ramp}=150\,\mathrm{ms}$. Since typically
   the chemical potentials $\mu_0$ of our condensates in traps with
   frequencies ${\overline\omega}_{trap}/2\pi\approx 20-70\,\mathrm{Hz}$ lie
between $50\,\mathrm{Hz}$
   and $200\,\mathrm{Hz}$, the adiabaticity
condition~\cite{band02}
   $t_{ramp}>h/\mu_0$ was satisfied even for small
   chemical potentials. After that, {\em only} the magnetic trap was
switched
   off and the condensate was imaged after a variable time
   $t_{exp}$ of free expansion inside the optical lattice. The effect
of a deep lattice on the
   free expansion is clearly evident in  the condensate images of
Fig.~\ref{Fig1:withandwithout} and in the measurements of the
condensate dimensions along ($\rho_{||}$) and perpendicular to the
lattice direction ($\rho_{\perp}$) shown in
Fig.~\ref{Fig2:expansion} ($\rho_{||,\perp}$ denoting the $e^{-1}$
half-width of a Gaussian fit to the density profile):
    Whilst in the lattice
   direction the condensate does not expand at all, the expansion
   in the direction perpendicular to the lattice is noticeably
   enhanced with respect to
   the expansion of the condensate in the absence of the optical
   lattice. The lack of expansion in the lattice direction
reflects the fact that the condensate
   has effectively been split up into several smaller condensates
confined in the individual lattice wells, whereas the
   enhanced expansion in the perpendicular direction is explained
by the increase in the chemical
   potential when the lattice is  ramped up.

   Taking into
account the periodic potential,
    Pedri {\em et
   al.} have calculated~\cite{pedri01} that in the limit of large
lattice depth ($U_{0}\gg E_{rec}$) and assuming that the
Thomas-Fermi approximation is
   always valid in the radial direction,
   the (local) chemical potential $\mu_{k}$ of well $k$ is given by
   \begin{equation}
   \mu_k= \frac{1}{2}m\omega_{||}^2d^2(k_m^2-k^2),
   \end{equation}
   where $\omega_{||}$ is the magnetic trap frequency in the  lattice
direction (similarly, $\omega_{\perp}$ will indicate the trap
frequency along the (observable) orthogonal direction).  The well
number $k$ ranges from $0$ to the
   maximum $k_m$ given by
   \begin{equation}
   k_m=\left(\frac{2\hbar \overline{\omega}_{trap}}{m\omega_{||}^2 d^2}
   \right)^{1/2}\left(\frac{15}{8\sqrt{\pi}}N\frac{a_s}{a_{ho}}
   \frac{d}{\sigma_{||}}\right)^{1/5},
   \end{equation}
   where  $N$ is the total number of condensate atoms,
   $a_{ho}=\sqrt{\hbar/(m\overline{\omega}})$ the harmonic
oscillator length, $a_s$ is the $s$-wave scattering length and
$\sigma_{||}$ is the width along the lattice direction of a
Gaussian wavepacket in a single lattice well.
    We calculated $\sigma_{||}$
   by using a variational ansatz with a Gaussian
   wavefunction~\cite{cristiani02}. The
total number of wells is given by
   $n=2k_m+1$. For the central well with $k=0$, the (local) chemical
   potential is
   \begin{equation}
   \label{eqn:pedri}
   \mu_{k=0} = \left(\frac{\pi}{2}\right)^{1/5}
   \left(\frac{U_0}{E_{rec}}\right)^{1/10}\mu_0,
   \end{equation}
   where $\mu_0$ is the chemical potential in the absence of the
   lattice. We have checked that considering only the central well
   for the perpendicular width reproduces to within a small error
   ($<1\%$)
   the result of a fit to the envelope of all lattice wells.

\begin{figure}
   \centering\begin{center}\mbox{\epsfxsize 2.8 in \epsfbox{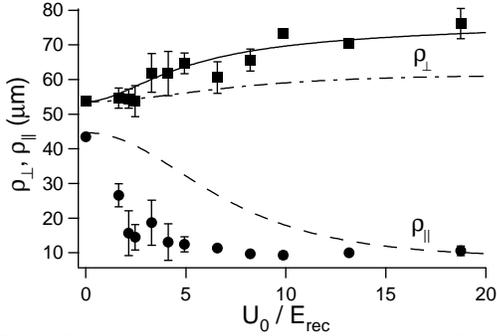}}
   \caption{Dependence on the lattice depth of the size of a
condensate released from a trap with
$\overline{\omega}_{trap}/2\pi=25\,\mathrm{Hz}$
($\omega_{\perp}/\omega_{||}=\sqrt{2}/1$) after a time-of-flight
of
   $22\,\mathrm{ms}$ for a vertical lattice. For $\rho_{\perp}$ the
solid line is the result of the numerical integration of the
scaling equations (see text) taking into account the variation of
the chemical potential $\mu_{k=0}$ with the lattice
    depth, whereas the dot-dashed line was calculated assuming a fixed
$\mu_0$. In the lattice direction (circles), the experimentally
measured expansion for $\rho_{||}$ lies below the theoretical
prediction. The error bars indicate the standard deviations of the
averaged data points.}\label{Fig3:vardepth}
   \end{center}\end{figure}

  We modeled the expansion of the condensate in the presence of the
1D lattice using the equations
   derived in~\cite{castin96,kagan96} with a slight
   modification along the lines of Ref.~\cite{kramer02}: In the ansatz leading to the differential
   equation of~\cite{castin96}
for the scaling
   factor $\lambda_{||}(t=t_{exp})= \rho_{||}(t=t_{exp})/\rho_{||}(t=0)$ along the lattice direction
   we replaced the atomic mass $m$
   by the effective mass $m^*(U_0)$, introduced in~\cite{kramer02}
   in analogy with a solid state physics approach, as derived from a
band structure
   calculation for a periodic potential of depth $U_0$. As can be seen
in Figs.~\ref{Fig2:expansion} and~\ref{Fig3:vardepth}, taking into
account the variation of $\mu_{k=0}$ with $U_0$ this model
reproduces well our experimental data for the perpendicular
expansion of the condensate (the theoretical plots are corrected
for the $5\,\mathrm{\mu m}$ resolution of our imaging system).
    For the lattice direction, this
   approach gives the correct result for deep lattices
   ($U_0\gg E_{rec}$) and reproduces reasonably well the
   qualitative behaviour in the intermediate regime
   ($0<U_0<5\,E_{rec}$). We find, however, that the experimental
expansion along the lattice direction is considerably less
   than the theoretical prediction. We have checked that taking into
account the finite momentum spread of the condensate when
   calculating the effective mass only leads to a correction on the
percent level and thus cannot explain the deviation of our
   experimental data from the numerical calculations neglecting
   mean field corrections. This might indicate that there are effects
    such as self-trapping~\cite{trombettoni01} due
   to the mean-field interaction that further reduce the expansion in
the lattice direction.

 For the perpendicular direction we have
   included in Fig.~\ref{Fig3:vardepth} the results of a numerical simulation {\em without} taking into account the
   variation of the chemical potential with the lattice depth, {\it i.e}
   using $\mu_{0}$ instead of $\mu_{k=0}$ of Eqn.~\ref{chemical}.
   Clearly, the resulting increase in perpendicular size is much
   less than what we observe experimentally and is related to the
   different re-distribution of mean-field energy when the degree
   of freedom in the lattice direction is frozen out. We have
   checked that the variation in lattice depth during the
   time-of-flight of up to $23\,\mathrm{ms}$ in the Gaussian beam
   profile (during which time the condensate drops by up to
   $2.6\,\mathrm{mm}$) does not affect the final widths. In fact,
   the mean-field explosion (during which mean-field energy is
converted into kinetic energy) occurs during the first few
   milliseconds (equivalent to $< 100\,\mathrm{\mu m}$ of
   free fall), after which the confinement along the lattice
   direction is almost perfect as long as the local lattice depth
   is larger than a few recoil energies.

\begin{figure}
   \centering\begin{center}\mbox{\epsfxsize 2.6 in \epsfbox{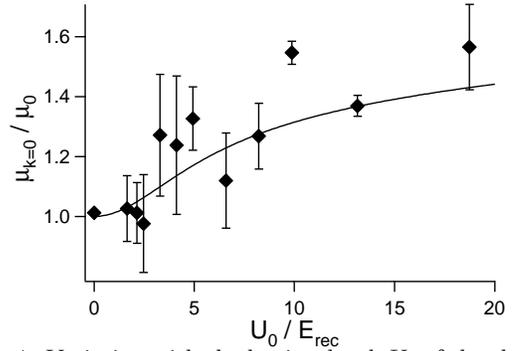}}
   \caption{Variation with the
   lattice depth $U_0$ of the chemical potential $\mu_{k=0}$ (in
units of $\mu_0$ in the absence of the lattice)  calculated from
the data of Fig.~\ref{Fig3:vardepth}. The solid line is the
theoretical prediction of Pedri {\em et al}.}\label{Fig4:chempot}
   \end{center}\end{figure}

   We could also directly deduce from our data the dependence of the
   chemical potential on $U_0$ by calculating
    $\mu_{k=0}$ from the initial perpendicular size
   $\rho_{\perp}(t=0)$ of the condensate in the presence of the lattice
inferred from
   the size $\rho_{\perp}(t=t_{exp})$ measured after an expansion time
$t_{exp}$ and the ratio
   $\lambda_{\perp}(t=t_{exp})/\lambda_{\perp}(t=0)$ of the scale factor
   $\lambda_{\perp}(t)$. The chemical potential
   was then given by
   \begin{equation}
   \mu_{k=0}=\frac12 m\omega_{\perp}^2 \rho_{\perp}^2(t=0).
   \label{chemical}
   \end{equation}
  In Fig.~\ref{Fig4:chempot} we plot the ratio
$\mu_{k=0}/\mu_0$ between the chemical potential
   as measured from the expansion of the condensate in the optical
   lattice and the chemical potential $\mu_0$ in the magnetic trap
alone. Agreement
   with Eqn.~\ref{eqn:pedri} is very good.

   \begin{figure}
   \centering\begin{center}\mbox{\epsfxsize 2.6 in \epsfbox{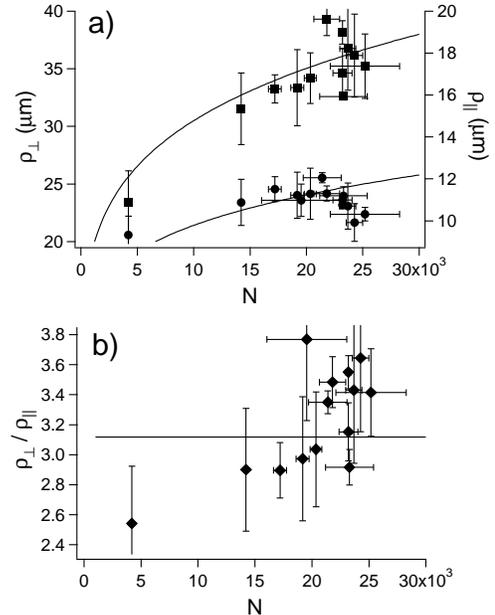}}
   \caption{Variation of the condensate widths (a) along
($\rho_{||}$, circles) and perpendicular to the lattice direction
    ($\rho_{\perp}$, squares), and of the the aspect ratio (b) with
the number of
   atoms $N$. The solid lines are predictions based on Eqn.~\ref{chemical} for
the chemical potential $\mu_{k=0}$ in the
   presence of the lattice. In this experiment
$U_0/E_{rec}=20$, $\overline{\omega}_{trap}/2\pi=12\,\mathrm{Hz}$
with $\omega_{\perp}/\omega_{||}=\sqrt{2}/1$. The error bars
represent standard deviations for the averaged data points at
fixed final RF cut.}\label{Fig5:widthsN}
   \end{center}\end{figure}

   We have also checked the dependence of $\mu_{k=0}$ on the number
   of atoms $N$ in the condensate. As $N$ is varied by cutting
into the (almost pure) condensate during the
   final stage of the RF evaporation, for a deep lattice the size of
the
   condensate after a time-of-flight increases considerably in the
perpendicular
   direction when $N$ increases, as expected from the variation in
chemical potential (see
   Fig.~\ref{Fig5:widthsN}~(a)). In the lattice direction,
    however, the increase in size shows a weaker dependence on $N$
than expected. For $N>2\times 10^4$, $\rho_{||}$ even
    starts to decrease with increasing condensate number.
Consequently, the aspect ratio $\rho_{\perp}/\rho_{||}$ increases
with $N$ (see Fig.~\ref{Fig5:widthsN}) rather than remaining
constant as theoretically predicted.
   This behaviour suggests that the condensate width in the lattice
direction does not fully reflect the change in chemical potential.
When $N$ gets large, mean-field effects might become important in
the dynamics of the condensate expansion, as mentioned above,
   and non-linear effects like self-trapping might reduce the observed
width $\rho_{||}$.

   Finally, we note here that in the limit of large lattice depths, our
   experiments effectively realize an adiabatic transformation
   between a 3D condensate and an array of 2D condensates. The
condition $\mu_{3D}<\hbar \omega_{lat}$ of Ref.~\cite{gorlitz01}
    (where $\omega_{lat}= 2(E_{rec}/\hbar)\sqrt{U_0/E_{rec}}$ is the
   harmonic approximation for the oscillation frequency in a lattice
well) for
   the condensates in each well to be in the 2D limit is always
satisfied for the small number of atoms
   in a single well ($\approx 10^3$) present in our
experiment. For an
   array of 2D condensates obtained by creating the condensate in the
combined potential
    of the harmonic trap and the lattice, Burger {\em et
al.}~\cite{burger02} have shown that
   in the case of their cigar-shaped condensate (with the long axis
   along the lattice direction), the transition temperature $T_c^{2D}$
in the
   presence of the lattice is significantly lower than $T_c^{3D}$ in
the 3D
   case (i.e. in the magnetic trap without the lattice).
   Calculating the critical temperature $T_c^{2D}$ along the same lines
   for our system, we find that $T_c^{2D}\approx T_c^{3D}$ due to the
larger number of atoms per
   lattice site in our geometry, and
   hence we expect no significant change in the condensate fraction in
the
   presence of the lattice. In fact, experimentally we even find a
consistently
   larger condensate fraction after ramping up the lattice. This
   result indicates that, with an appropriate choice of parameters, a
1D optical lattice could be used to
   investigate adiabatic transformations between 3D and 2D
   condensates which could, e.g., be exploited to create
   condensates from thermal clouds by changing the dimensionality
   of the system, similarly to the change in the shape of the potentials in Refs~\cite{stamperkurn99,pinkse97} for other
   geometries.

   In summary, we have studied the free expansion of a
   Bose-Einstein condensate in a one-dimensional optical lattice.
   When increasing the lattice depth, an increase in the chemical potential of the condensate
   was measured, in agreement with a recent
   theoretical calculation~\cite{pedri01}. The expansion along the
lattice direction, though,
    seems to be strongly reduced in the regime of intermediate lattice
depth, necessitating a more
    thorough theoretical treatment taking into account mean-field
effects in the dynamics after the magnetic trap is
    switched off. Furthermore, the possibility of
   adiabatically changing the dimensionality between a 3D trap and an
   array of 2D traps should, with appropriate experimental
   parameters, represent a way to reach quantum degeneracy by
   adding a periodic potential to a harmonic trap.  The improved
   characterization  of the
   behaviour of condensates in optical lattices is important
 with regard to adiabaticity criteria when a
condensate is
   transferred into a periodic potential that could, e.g., be used
   to implement a neutral atom quantum computer.

   The authors would like to thank S. Stringari, Y. Castin and W.D.
Phillips for stimulating
   discussions. This work was supported by the MURST (PRIN2000
   Initiative), the INFM (Progetto di Ricerca Avanzata
   `Photonmatter'), and by the the EU through the Cold Quantum
   Gases Network, Contract No. HPRN-CT-2000-00125. O.M. gratefully
   acknowledges a Marie Curie Fellowship from the EU within the IHP
   Programme.

   \end{document}